\documentstyle[prl,aps,epsf,multicol]{revtex}
\begin{document}
\sloppy
\draft
\title
{Dynamo effect beyond the kinematic approximation: suppression of linear
instabilities by non-linear feedback}
\author
{Abhik Basu\cite{byemail}}
\address{Poornaprajna Institute of Scientific Research,
Bangalore, India,\\and\\
Centre for Condensed Matter Theory,
Department of Physics, Indian Institute of Science,
Bangalore 560012, India.}
\maketitle
\begin{abstract}
The turbulent dynamo effect, which describes the generation of magnetic fields
in astrophysical objects, is described by the dynamo equation. This, in the
kinematic (linear) approximation gives an {\em unbounded exponential} growth of
the long wavelength part of the magnetic fields. Here we, in a systematic
diagrammatic, perturbation theory,
show how non-linear effects suppress the linear instability and bring down the
growth rate to zero in the large time limit. We work with diffferent background
velicity spectrum and initial magnetic field correlations. Our results indicate
the robustness and very general nature of dynamo growth: 
It is qualitatively independent of
the background velicty and intial magnetic field spectra. 
We also argue that our results can
be justified within the framework of the first order smoothing approximation,
as applicable for the full non-linear problem. We discuss our results from the
view points of renormalisation group analysis.

\end{abstract}

\section{Introduction}
Magnetic fields are ubiquitous. All astrophysical objects are known to have
magnetic fields of different magnitudes,e.g., 1 gauss at the stellar scale
to $10^{-6}$ gauss at the galactic scale \cite{moff}. The origin of this
field ({\em primordial field}) is not very clear - there are several 
competing theories which attempt to describe this \cite{kron}. However,
a finite magnetic field in any physical system undergoes a temporal 
decay due to
the finite conductivities of the medium. So, for steady magnetic fields 
to remain in astrophysical bodies, there has to be regeneration of the
magnetic fields which takes place in the form of dynamo process
\cite{moff,abhik}. Typically astrophysical bodies are thought to have
{\em fast} dynamo operating within themselves, resulting into 
exponential growth of the magnetic fields. This mechanism requires 
a turbulent velocity background \cite{moff} [though non-turbulent velocity 
fields too can make a seed (initial) magnetic field to grow (for 
details see \cite{abhik}) we will not consider such cases here].
Since the dynamo equations, in the linear approximation
(see below) gives unbounded exponentially growing solutions for the long
wavelength (large scale) part of the magnetic fields, it is linearly
unstable in the low wavenumber limit. An intriguing question, that arises very 
naturally is, what happens at a later time, i.e., whether magnetic 
fields continue to grow even after a long time. This obviously does 
not happen as we do not see ever growing magnetic fields in the core
of the earth or in the sun. For example, geomagnetic fields ($\sim$ 1 gauss)
are known to be stable for about $10^6$ years \cite{moff}.
Secondly, continuously growing magnetic 
fields violate energy conservation. In other words, if the dynamo 
equation correctly describes the problem, then its physically realisable
solutions must not be linealy unstable in the long time limit. So there must 
be a counter mechanism to stabilise it, which we investigate here. 

There have
been numerous works in this field in the past by many groups. For examples
Pouquet, Frisch and L\'{e}orat \cite{po} in a eddy damped quasi-normal Markovian
approximation studied the connections between the dynamo process and the
inverse cascade of magnetic and kinetic energies. Moffatt \cite{moff1}, by
linearising the equations of motion of three-dimensional ($3d$)
magnetohydrodynamics (MHD) examined the back reactions due to the
Lorentz force for magnetic Prandtl number $P_m\gg 1$. Vainshtein and Cattaneo
\cite{vain} discussed several nonlinear restrictions on the generations of
magnetic fields. Field {\em et al} \cite{field} discussed nonlinear
$\alpha$-effects within a two-scale approach. Rogachevskii and Kleeorin
\cite{roga} studied the effects of an anisotropic background turbulence on the
dynamo process. Brandenburg examined non-linear $\alpha$-effects in numerical
simulation of helical MHD turbulence \cite{brand}. He particularly examined the
dependences of dynamo growth and the saturation field on $P_m$. Bhattacharjee
and Yuan \cite{bhatta} studied the problem in a two-scale approach by
linearising the equations of motion. However these issues are not yet fully
closed. We examine the following questions: i)instead of a two-scale approach
(which is rather adhoc) whether we can employ a diagrammatic perturbation
theory, which has been highly successful in the context of critical dynamics
\cite{halparin}, driven systems \cite{fns} etc., can be easily extended to
higher orders in perturbation expansion and provides natural connections with
standard renormalisation group framework, and ii)if a turbulent
background {\footnote{By a turbulent background we do not mean any kind of
fluctuating state but
a fluctuating state with K41 spectrum for the kinetic and magnetic energies and
cascades of appropriate quantities; if there is no mean magnetic field then the
energy spectra is expected to be K41-type - see Ref.\cite{abprl}.} }
is essential for the dynamo mechanism. To
put it differently we ask if dynamo process can take place with velocity fields
with arbitrary statistics.
We explicitly demonstrate that the nonlinear feedback of the magnetic fields on
the velocity fields in the form of the Lorentz force stabilises the
instability.  We show it for a very general velocity and initial magnetic field
correlations - thus our results demonstrate the very generality of the dynamo
process. The plan of the rest of the paper is as follows:
In Section \ref{dyn} we discuss the general dynamo mechanism
within the standard linear approximation. In Sec. \ref{halt} we
show that one needs to go beyond the linear approximation, i.e., include the
non-linear effects to see eventual saturation of magnetic field growth.
In Sec. \ref{summ} we conclude. 

\section{Dynamo growth: the linear approximation}
\label{dyn}
In the kinematic approximation \cite{moff,abhik1}, i.e., in the early time when 
magnetic energy
is much smaller than the kinetic energy ($\int u^2 d^3r >> \int b^2
d^3 r$), where ${\bf u(r},t)$ and ${\bf b(r},t)$ are the velocity and magnetic
fields respectively, the Lorentz force term of the Navier Stokes 
equation is neglected. In that weak magnetic field limit, which is 
reasonable at an early time, the time evolution problem of the magnetic
fields is a linear problem as the Induction equation \cite{jac}
is linear in magnetic fields $\bf b$:
\begin{equation}
{\partial {\bf b}\over \partial t}=\nabla\times ({\bf u\times b})+\mu
\nabla^2 {\bf b},
\label{indeq}
\end{equation}
where $\mu$ is the magnetic viscosity.  The velocity field is governed by the
Navier-Stokes equation \cite{land} (dropping the Lorentz force)
\begin{equation}
{\partial{\bf u}\over\partial t}+{\bf u.\nabla u}=-{\nabla p\over\rho}
+\nu\nabla^2 {\bf u}+{\bf f}.
\end{equation}
Here $\nu$ is the fluid viscosity,  $\bf f$ an external forcing function,
$p$ the pressure and $\rho$ the density of the fluid.
We take $\bf f$ to be a zero mean, Gaussian stochastic force with a
specified variance (see below).

In a two-scale \cite{moff} approach one can then write an {\em effective} 
equation for ${\bf B}$, the long-wavelength part of the magnetic fields 
\cite{moff}:
\begin{equation}
{\partial {\bf B}\over \partial t}=\nabla\times ({\bf U\times B})+
\nabla\times {\bf E}+\mu\nabla^2 {\bf B},
\end{equation}
where the {\em Electromotive force} $\bf E=\langle v\times b\rangle$. 
$\bf U$ is the large scale component of the velocity field $\bf u$. An 
{\em Operator Product Expansion} (OPE) is shown to hold \cite{abhik1} 
which provides a gradient expansion in terms of $\bf B$ for the product  
$\bf E=\langle u\times b\rangle$ \cite{moff}
\begin{equation}
E_i=\alpha_{ij}B_{j}+\beta_{ijk}{\partial B_{j}\over \partial x_k}+... .
\label{ope}
\end{equation}
For homogenous and isotropic flows ($\alpha_{ij}=\alpha\delta_{ij}$) 
Eq.(\ref{ope}) gives,
\begin{equation}
{\partial {\bf B}\over\partial t}=\nabla\times ({\bf U\times B})
+\alpha\nabla\times {\bf B}+\mu\nabla^2{\bf B},
\label{dyna}
\end{equation}
which is the standard turbulent dynamo equation. Here $\mu$ now is the
{\em effective} magnetic viscosity which includes turbulent diffusion,
represented by $\beta_{ijk}$ in Eq.(\ref{ope}). $\alpha$  depends
upon the statistics of the velocity field (or, equivalently, the correlations of
$\bf f$). Retaining only the $\alpha$
-term and dropping all others from the RHS of Eq.(\ref{dyna}), the
equations for the cartesian components of $\bf B$ become (we neglect the
dissipative terms proportional to $k^2$ as we are interested only in the long
wavelength properties)\\
\[
{d\over dt}
\pmatrix{
B_x({\bf k},t) \cr B_y({\bf k},t) \cr B_z({\bf k},t)}
=i\alpha\pmatrix{
0 & -k_z & k_y \cr
k_z & 0 & -k_x \cr
-k_y & k_x & 0}
\pmatrix{
B_x({\bf k},t) \cr B_y({\bf k},t) \cr B_z({\bf k},t)}.
\]
The eigenvalues of the matrix is {$\lambda=\pm ik,\,0$}. Thus depending on the
sign of the product $\alpha k$, one mode grows and the other decays. 
The third mode
stays constant in time. Since growth rate is proportional to 
$|k|$ and dissipation
proportional to $k^2$, large scale fields continue to grow leading to 
long wavelength
instability. Thus in the long time limit effectively only the growing mode
remains. Since the cartesian components of $\bf B$ are just linear 
combinations of the eigenmodes of the matrix above, they also grow 
exponentially. Growth rate $\alpha$ is a pseudo-scalar quantity, i.e., under
parity transformation $\bf r\rightarrow -r$, $\alpha\rightarrow -\alpha$ 
\cite{moff,abhik1}. Since $\alpha$ depends upond the statistical properties of
the velocity field, its statistics  should not be parity invariant.
This can happen in a rotating frame, where the angular
velocity explicitly breaks parity.

\section{Formulation of the dynamo problem in a rotating frame}
\label{rot}
The NS and the Induction equation in the inertial (lab)
frame in $({\bf k}, t)$ space become
\begin{equation}
{\partial u_i({\bf k},t)\over\partial t}+{1\over 2}P_{ijp}({\bf k})
\sum_q u_j({\bf q},t)
u_p({\bf k-q},t)={1\over 2}P_{ijp}({\bf k})\sum_q b_j({\bf q},t)
b_p({\bf k-q},t)+ \nu\nabla^2 u_i +f_i({\bf k},t),\\
\label{nsk}
\end{equation}
\begin{equation}
{\partial b_i({\bf k},t)\over\partial t}=\tilde{P}_{ijp}({\bf k})
\sum_q u_j({\bf q},t) b_p({\bf k-q},t)+\mu\nabla^2 b_i.
\label{indk}
\end{equation}
Here, $u_i({\bf k},t)$ and $b_i({\bf k},t)$ are the fourier transforms of
$u_i({\bf r},t)$ and $b_i({\bf r},t)$ respectively, $P_{ijp}({\bf k})=P_{ij}
({\bf k})k_p+P_{ip}({\bf k})k_j,\,\tilde{P}_{ijp}({\bf k})=P_{ij}({\bf k})k_p
-P_{ip}({\bf k})k_j,\,P_{ij}$ is the projection operator, which appears due to
the divergence-free conditions on the velocity and magnetic fields. The
Eqs.(\ref{nsk}) and (\ref{indk}) are to be supplimented by appropriate
correlations for $f_i$ and intial conditions on $b_i$. We choose $f_i({\bf
k},t)$ and $b_i({\bf k},t=0)$ to be zero mean Gaussian distributed with the
following variances:
\begin{equation}
\langle f_i({\bf k},t)f_j(-{\bf k},0)\rangle=P_{ij}D_1(k)\delta(t),\\
\label{varf}
\end{equation}
\begin{equation}
\langle b_i({\bf k},t)b_j(-{\bf k},0)\rangle=P_{ij}D_2(k)\delta(t).
\label{varb}
\end{equation}
$D_1$ and $D_2$ are some functions of $k$ (to be specified later).

In a rotating frame with a rotation velocity ${\bf \Omega}=\Omega\bf \hat{z}$ 
the Eqs.(\ref{nsk}) and (\ref{indk}) take the form
\begin{equation}
{\partial u_i({\bf k},t)\over\partial t}+2({\bf \Omega\times u})_i+
{1\over 2}P_{ijp}({\bf k}) \sum_q u_j({\bf q},t)
u_p({\bf k-q},t)={1\over 2}P_{ijp}({\bf k})\sum_q b_j({\bf q},t)
b_p({\bf k-q},t)+ \nu\nabla^2 u_i +f_i({\bf k},t),\\
\label{nskr}
\end{equation}
\begin{equation}
{\partial b_i({\bf k},t)\over\partial t}+({\bf \Omega\times b})_i
=\tilde{P}_{ijp}({\bf k})
\sum_q u_j({\bf q},t) b_p({\bf k-q},t)+\mu\nabla^2 b_i.
\label{indkr}
\end{equation}
$\bf \Omega\times u$ is the coriolis force. The centrifugal force
$\bf \Omega\times(\Omega\times r)$ is put in as a part of the {\em effective
pressure}=$p+{1\over 2}|{\bf \Omega\times r}|^2$ which does not contribute to
the dynamics for incompressible flows.
The correlations Eqs.(\ref{varf}) and (\ref{varb}) do not
change in the rotating frame. The bare propagators $G_u$ and $G_b$ 
of $u_i$ and $b_i$ are\\
\[G_u=\pmatrix{{{i\omega+\nu k^2}\over (i\omega +\nu k^2)^2+4\Omega^2}& 
-{2\Omega\over (i\omega+\nu k^2)^2+4\Omega^2} 
& 0 \cr
{2\Omega\over (i\omega+\nu k^2)^2+4\Omega^2} & {{i\omega +\nu k^2} \over
(i\omega +\nu k^2)^2+4\Omega^2}& 0 \cr
0 & 0 & {1\over i\omega+\nu k^2}} \;\&\;
G_b=\pmatrix{{{i\omega+\mu k^2}\over (i\omega +\nu k^2)^2+\Omega^2} & 
-{\Omega\over (i\omega+\mu k^2)^2+\Omega^2}
& 0 \cr
{\Omega\over (i\omega+\mu k^2)^2+\Omega^2} & {{i\omega +\mu k^2} \over 
(i\omega +\nu k^2)^2+\Omega^2}& 0 \cr
0 & 0 & {1\over i\omega+\mu k^2}}
\]
\\
such that $\bf u= G_u\,f$ and ${\bf b}(t)={\bf G_b\,b}(t=0)$ where 
\[{\bf u}=\pmatrix{u_x\cr u_y\cr u_z},\,{\bf b}=\pmatrix{b_x\cr b_y\cr b_z}.
\]
\\
We get for the correlation function $\langle u_i({\bf k},\omega) u_j({\bf
-k},-\omega)\rangle \equiv \langle {\bf u}{\bf u}^T\rangle$ (
${\bf u}^T$ is the transpose of $\bf u$), when $i\neq j$ there
are terms proportinal to $\Omega$ (for $i=j$, there are no $O(\Omega)$ terms). 
Under
parity transformation these $i\neq j$ terms change sign. Thus the $i\neq j$
terms have both even parity and odd parity parts. Similarly in $\langle b_i b_j
\rangle,\, {\rm for}\,i\neq j$ have both odd and even parity parts. 
A direct consequence
of these parity breaking parts is that fluid helicity ($\equiv \int_{\bf x}
\langle{\bf u.(\nabla\times u)}\rangle$) is non-zero:
$\int_{\bf x}\langle{\bf u.(\nabla\times u)}\rangle\propto \Omega.$
In the same way $\int_{\bf x}\langle{\bf b.(\nabla\times b)}\rangle$ 
is non-zero and proportional to $\Omega$.
Both $\int_{\bf x}\langle{\bf u.(\nabla\times u)}\rangle$ and
$\int_{\bf x}\langle{\bf b.(\nabla\times b)}\rangle$ have same signs - a fact
of great importance for the results discussed here. Notice that $G_{zz}^{u,b}$
are different from $G_{xx,yy}^{u,b}$ - this is just the consequence of the
fact that $\Omega$ distinguishes the $z$-direction from others, making the
system anisotropic. However for frequencies $\omega >>\Omega$ or length scales
$k^z>>\Omega$ isotropy is restored. In that regime the role of
the global rotation is only to introduce parity breaking contributions 
proportional to $\Omega$ to
$\langle u_iu_j\rangle$ and $\langle b_ib_j\rangle$ for $i\neq j$. These can
be {\em modeled} by introducing parity breaking parts in Eqs.(\ref{varf}) and
(\ref{varb})
\begin{eqnarray}
\langle f_i({\bf k},t)f_j(-{\bf k},0)\rangle&=&P_{ij}D_1(k)\delta(t)+
2i\epsilon_{ijp}k_p\tilde{D}_1(k)\delta(t),\\
\label{varfi}
\langle b_i({\bf k},t)b_j(-{\bf k},0)\rangle&=&P_{ij}D_2(k)\delta(t)+
2i\epsilon_{ijp}k_p\tilde{D}_2(k)\delta(t),
\label{varbi}
\end{eqnarray}
in conjunction with the inertial frame Eqs.(\ref{nsk}) and (\ref{indk}). The
parity breaking parts in the noise correlations or intial conditions ensure
that $\int_{\bf x}\langle{\bf u.(\nabla\times u)}\rangle$ and 
$\int_{\bf x}\langle{\bf b.(\nabla\times b)}\rangle$ are non-zero as is the 
case with Eqs.(\ref{nskr}) and Eqs.(\ref{indkr}) along with Eqs.(\ref{varf})
and (\ref{varb}). What is the relative sign between $\tilde{D}_1$ and 
$\tilde{D}_2$? Since $\int_{\bf x}\langle{\bf u.(\nabla\times u)}\rangle$
and $\int_{\bf x}\langle{\bf b.(\nabla\times b)}\rangle$ are proportional to
$\tilde{D}_1 \;{\rm and}\; \tilde{D}_2$ respectively, and they have same signs,
$\tilde{D}_1$ and $\tilde{D}_2$ must have same signs.

\subsection{Calculation of $\alpha$ in the kinematic approximation}

In the kinematic approximation neglecting the Lorentz force term of the 
Navier-Stokes equation, the time evolution of the magnetic 
fields follows from the linear Induction Equation (\ref{indeq}). 
We assume, for convenience of calculaions,
that the velocity field ($\bf u$) statistics has reached a steady state.
This is acceptable as long as the loss due to the transfer of kinetic energy by 
the dynamo process is compensated by the external drive. In the 
kinematic (i.e., linear) approximation, we work with the Eqs.(\ref{nsk})
(without the Lorentz force) and (\ref{indk}).
We choose $f_l({\bf k},t)$ to be a zero-mean,
 Gaussian random field with correlations
\begin{equation}
\langle f_l({\bf k},t)f_m({\bf k},0)\rangle=2P_{lm}D_1(k)\delta(t)+
2i\epsilon_{lmn}\tilde{D}_1(k)k_n\delta(t).
\end{equation}
Our intial conditions for the magnetic fields are
\begin{equation}
\langle b_{\alpha}({\bf k},t=0)b_{\beta}(-{\bf k},t=0)\rangle=P_{\alpha\beta}
2D_2(k)+2i\epsilon_{\alpha\beta\gamma}k_{\gamma}\tilde{D}_2(k),
\end{equation}
Since we are interested to investigate the dynamo process with arbitrary
statistics for the velocity and magnetic fields we work with arbitrary $D_1(k),
\tilde{D}_1(k),D_2(k)$ and $\tilde{D}_2(k)$. For K41 spectra, we require
\cite{yakhot} $D_1(k)=D_1k^{-3},\tilde{D}_1(k)=\tilde{D}_1k^{-4},D_2(k)=D_2
k^{-5/3}$ and $\tilde{D}_2(k)=k^{-8/3}$. These choices ensure that under
spatial rescaling ${\bf x}\rightarrow l{\bf x}$, ${\bf v,b}\rightarrow l^{1/3}
\{ {\bf v,b}\}$ which is the Kolmogorov scaling. 
Starting with such a correlations
ensures that only the amplitudes of the magnetic field correlations grow in time
due to the dynamo effects but the inertial range scale dependence does not
change.
However, this may not be the case always. In general, not only the amplitude 
grows in time due to dynamo actions, the scale dependence too can be anything
at $t=0$ and evolve in time. 
Note that both the force correlations in the Eq.(\ref{nsk}) and the
intial conditions on Eq.(\ref{indk}) have parts that are parity
breaking, in conformity with our previous discussions. 
We now calculate the $\alpha$-term
(at the tree level) in the kinematic approximation (which we call the
`direct' term - responsible for growth) below (see Fig.1a):
\begin{equation}
\langle ({\bf u\times b})_{\mu}\rangle_D=\langle
\int_q\epsilon_{\mu\beta\gamma}
u_{\beta}({\bf q},t)b_{\gamma}({\bf k-q},t)\rangle =\langle
\int_q \epsilon_{\alpha
\beta\gamma}u_{\beta}({\bf q},t)\epsilon_{\gamma\delta\lambda}i({\bf k-q})
_{\delta}u_{\eta}({\bf q_1},t_1)b_{\tau}({\bf k-q-q_1},t_1)G_0^b({\bf k-q},
t-t_1)\rangle
\end{equation}
which gives the $\alpha$-term:
\begin{equation}
\alpha_D B_{\alpha}({\bf k},t)=\int_q {i\tilde{D}_1(q) \over \nu q^2}
\epsilon_{\beta\eta\rho}q_{\rho}\epsilon_{\alpha\beta\gamma}(-i)q_{\delta}
b_{\rho}({\bf k},t=0)[{1\over q^2(\nu+\mu)}+{\exp(-2t\nu q^2)\over
q^2(\nu-\mu)}
\end{equation}
giving
$\alpha_D={2S_3\over 3}{1\over {\nu(\nu+\mu)}}\int_q 2{\tilde{D}(q)\over
(\nu+\mu)q^2}$
for large $t$. Thus self-consistently,
$\alpha_D=-{2S_3\over 3}\int_q 2{\tilde{D}_1(q)\over\nu[|\alpha_Dq|-
(\nu+\mu)q^3]}$.
The suffix $D$ refers to {\em growth} or the {\em direct} term, as opposed to
{\em feedback} which we discuss in the next Sec.\ref{halt}. The growth term is
proportional to $|k|$ and diffusive decay proportional to $k^2$. 
So large scale components grow
and small scale components decay. 
\begin{figure}[htb]
\centerline{
\epsfxsize=8cm
\epsffile{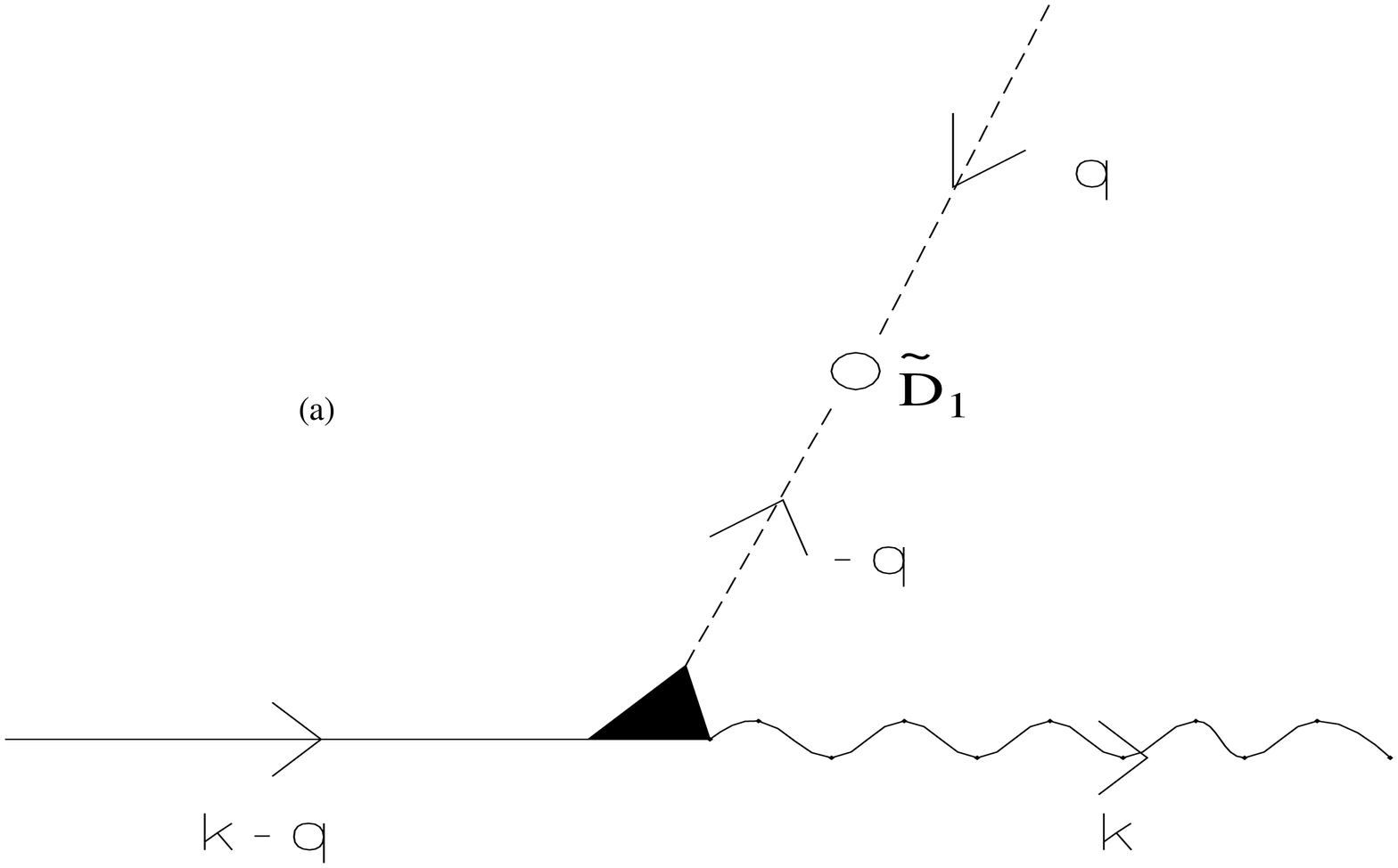}
\hfill
\epsfxsize=8cm
\epsffile{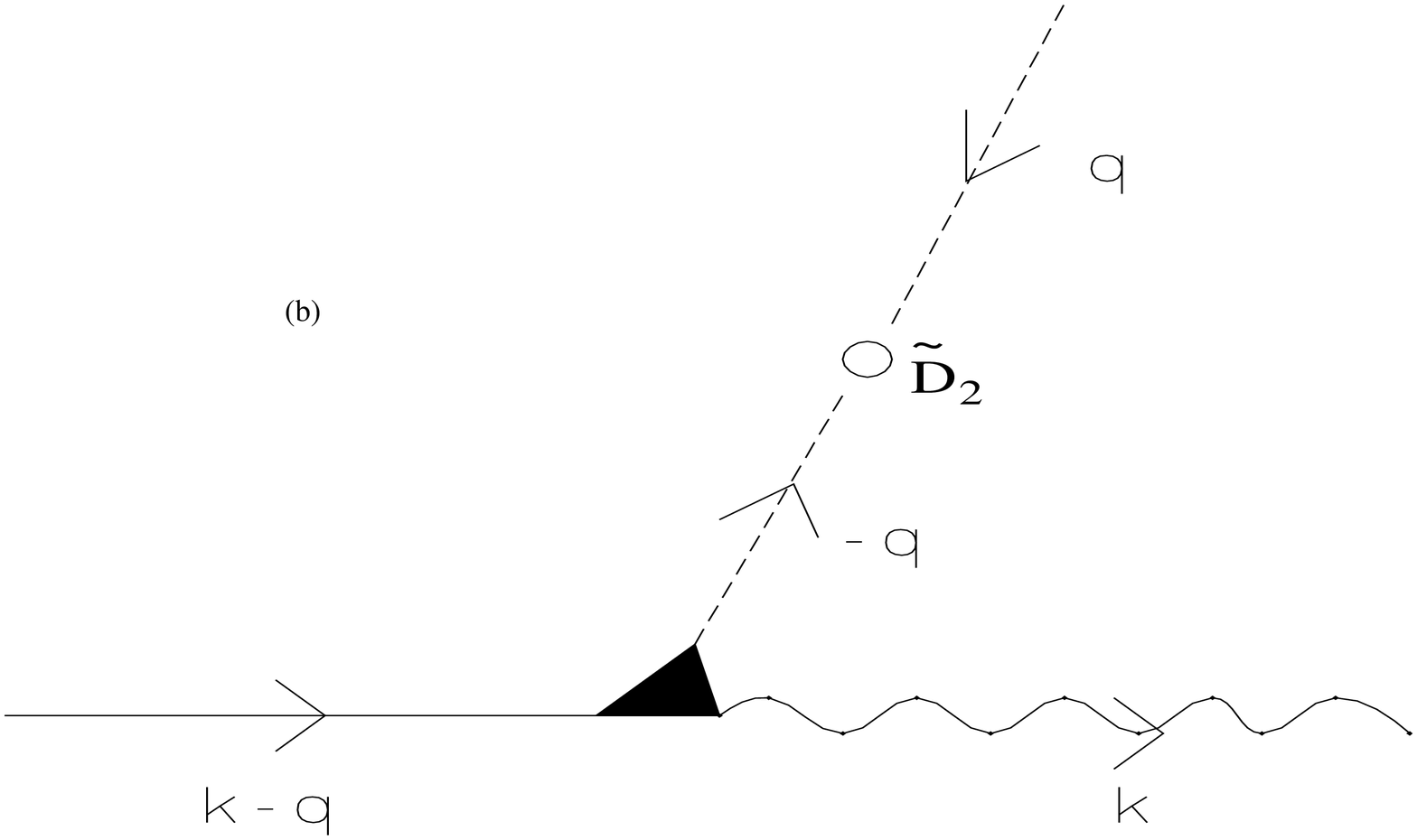}
}
\caption{Tree level diagrams for $<\bf u(q)\times b(k-q)>$. (a)A solid line
indicates a bare magnetic field response function, a broken line indicates
a bare velocity response funtion, a 'o' joined by two broken lines indicates
a bare velocity correlation function (proportional to $\tilde{D}_1$), 
a wavy line indicates a magnetic field,
a solid triangle indicates a $ub$ vertex. This contributes of $\alpha_D$ 
(b) A solid line
indicates a bare magnetic field response function, a broken line indicates
a bare velocity response funtion, a 'o' joined by two broken lines indicates
a bare magnetic field correlation function (proportional to $\tilde{D}_2$), 
a wavy line indicates a magnetic field,
a solid triangle indicates a $ub$ vertex. This contributes to $\alpha_F$.}
\label{diag}
\end{figure}

\subsection{Suppression of growth rate: Nonlinear feedback}
\label{halt}
When the magnetic fields become strong neglecting the feedback of the
magnetic fields in the form of the Lorentz force is no longer justified.
So we work with the {\em full} Eqs.(\ref{nsk}) and (\ref{indk}).
We here follow a diagrammatic perturbation approach.
In presence of the Lorentz force there is an additional contribution to 
$\alpha$ (Fig.1b). 
\begin{eqnarray}
\langle ({{\bf u\times b}_i})_F\rangle&=&\langle
\int_q \epsilon_{ijp}u_j({\bf q},t)
b_p({k-q},t)\rangle\\=\langle{i\over 2}\epsilon_{ijp}\int_q 
P_{jmn}(q)G_o^u(q,t-t_1)&&
b_m(q_1,t_1)b_n(q-q_1,t_1)G_o^b(k-q,t)b_p(k-q,t=0)\rangle
\end{eqnarray}
which gives ($F$ refers to feedback) 
\begin{equation}
\alpha B_i({\bf k},t)=i\epsilon_{ijp}\int_q P_{jmn}(q)e^{2\alpha_D|q|t-2\mu
q^2t}b_n({\bf k},t){-2i\tilde{D}_2(q)\epsilon_{mps}q_s\over 2\alpha_D |q|
-2\mu q^2}.
\end{equation}
This on simplification gives
\begin{equation}
\alpha_F={2S_3\over 3}{4\over 15}\int_q {\tilde{D}_2 (q,t)
\over \alpha_D|q|-2\mu q^2},
\end{equation}
where $\tilde{D}_2(q,t)=\exp[2\alpha_D|q|t-2\mu q^2t]\tilde{D}_2(q)$ is
a growing function of time for small wavenumbers. Thus $\alpha_F$ grows in
time.

Thus, at a late time $t$, when the non-linear feedback on the velocity field 
due to the Lorentz force is nolonger negligible, i.e., for a finite $\alpha_D$
and $\alpha_F$ we find self-consistently,
\begin{eqnarray}
\alpha_D&=&-{2S_3\over 3}\int {d^3q\over (2\pi)^3}{\tilde{D}_1(q)\over
{\nu [|(\alpha_D+\alpha_F)q|-(\nu+\mu)q^2]}},\\
\alpha_F&=&{2S_3\over 3}{4\over 15}\int {d^3q\over (2\pi)^3}{\tilde{D}_2(q,t)
\over {|(\alpha_D+\alpha_F)q|-2\mu q^2}}.
\end{eqnarray}
Thus net growth rate $\propto |(\alpha_D+\alpha_F)k|$ for the mode $B_i({\bf
k},t)$. 

Let us now consider various $k$ dependences of $\tilde{D}_1(k)$ and
$\tilde{D}_2(k)$. For the case when the background velocity field is driven by
the Navier-Stokes equation with a conserved noise (thermal noise) one requires
to have $D_1(k)=D_1k^2,\,\tilde{D_1}=\tilde{D}_1k$, giving $\langle v_i({\bf
k},t)v_i({\bf -k},t)\rangle=constant$. If we assume similar $k$-dependences for
$\langle b_i({\bf k},0)b_i({-\bf k},0)\rangle$ then we require $D_2(k)\sim$
constant and $\tilde{D}_2(k)={\tilde{D}_2\over k}$. These choices give
\begin{eqnarray}
\alpha_D&=&-{2S_3\over 3}\int {d^3q\over
(2\pi)^3}{\tilde{D}_1q\over\nu[|(\alpha_D+\alpha_F)q|-(\nu+\mu)q^2}],\\
\alpha_F&=&{2S_3\over 3}{4\over 15}\int {d^3q\over (2\pi)^3}{\tilde{D}_2(t)\over
q[|(\alpha_D+\alpha_F)q|-2\mu q^2}],
\label{alphathermal}
\end{eqnarray}
which remain finite even as the system size diverges.

Fully developed turbulent state, characterised by K41 energy spectra,  is
generated by $D_1(k)\sim k^{-3}$ and $\tilde{D}_1(k)=\tilde{D}_1k^{-4}$. In
addtion if we assume that the initial magnetic field correlations also have
$k41$ scaling then $D_2(k)\sim k^{-5/3}$ and
$\tilde{D}_2(k)=\tilde{D}_2k^{-8/3}$. If one starts with a K41-type initial
correlations for the magnetic fields, then at a later time the scale dependence
for the magnetic field correlations are likely to remain same; 
only the amplitudes grow.  Notice that the spectra diverge as
$k\rightarrow 0$, i.e.,  as the system size diverges. This is a typical
characteristic of fully developed turbulence. For such a system
self-consistently,

\begin{eqnarray}
\alpha_D&=&-{2S_3\over 3}\int {d^3q\over (2\pi)^3}{\tilde{D}_1q^{-4}\over\nu
[|(\alpha_D+\alpha_F)q|-(\nu+\mu)q^2]},\\
\alpha_F&=&{2S_3\over 3}{4\over 15}\int {d^3q\over (2\pi)^3}{\tilde{D}_2(t)
q^{-8/3}\over [|(\alpha_D+\alpha_F)q|-2\mu q^2]}.
\label{alphak41}
\end{eqnarray}
The notable difference between the expressions Eqs.(\ref{alphathermal}) and
Eqs.(\ref{alphak41}) for the $\alpha$ coefficients is that the $\alpha$
coefficients diverge with the system size when the energy spectra are singular
in the infra red limit (for fully developed turbulence).

In general, at early times (small $\alpha_F$), $\alpha_F$ increases 
exponentially
in time. For $t\stackrel{>}{\sim}{1\over \alpha_D}\ln\alpha_D$, $\alpha_D$ and
$\alpha_F$ are comparable. The growth rate of $\alpha_F$ comes down and
$\alpha_D$ decays. Since $\alpha_D$ and $\alpha_F$ have different signs, $
|(\alpha_D+\alpha_F)|\rightarrow 0$ as $t\rightarrow$ large. 
Thus the net growth rate comes down to zero. Hence, Eqs.(\ref{alphathermal})
and Eqs.(\ref{alphak41}) suggest that the early time growth and late time
saturation of magnetic fields take place for different kind of background
velicity correlations and initial magnetic field correlations. Thus dynamo
instability and its saturation are rather intrinsic properties of the $3d$MHD
equations in a rotating frame. Our results also suggest that these processes
may take place for varying magnetic Prandtl number $\mu/\nu$.
The above analysis crucially depends on the fact that $\alpha_D$ and 
$\alpha_D$ have opposite signs, which, in turn, imply that $\tilde{D}_1$
and $\tilde{D}_2$ have same signs. We have already seen
that in a physically realisable situation where parity is broken
entirely due to the global rotation,  $\tilde{D}_1$ and  $\tilde{D}_2$ indeed
have the same sign. Our results also suggest that the net growth rate is
proportional to the difference between the fluid and the magnetic helicities 
at any time $t$.

How can we understand our results in a simple way? To do that we resort to 
{\em first order smoothing approximation} \cite{moff,arnab}. In the kinematic 
limit, in this smoothing approximation to calculate 
$\langle \bf v\times b\rangle$ one considers only the Induction equation 
 as $\bf v$ is supposed to be given. However when one
goes beyond the kinematic approximation, one has to consider the Navier-Stokes
equation as well. Thus in the first-order smoothing approximation one writes
the equations for the fluctuations $\bf v$ and $\bf b$ as (to the first order)
\begin{equation}
{\partial {\bf b}\over\partial t}\approx \nabla \times ({\bf v\times 
\overline B}) + \nabla \times ({\bf \overline V \times b}),
\end{equation}
and
\begin{equation}
{\partial {\bf v}\over \partial t}\approx \ldots +({\bf \overline{B}.\nabla)
b},
\end{equation}
where the ellipsis refer to all other terms in the Navier-Stokes equation and 
$\bf \overline B$ and $\bf \overline V$ are the large scale ({\em mean field})
part of the velocity and magnetic fields \cite{moff,arnab}. With these
we can write
\begin{equation}
\langle {\bf v\times b}\rangle_i=\langle \epsilon_{ijp}v_jb_p\rangle=
\langle\epsilon_{ijp}v_jB_m
{\partial\over\partial x_m}v_p\rangle+\langle\epsilon_{ijp}b_pB_m
{\partial\over\partial x_m} b_j\rangle\equiv \alpha_{im}B_m+\ldots
\end{equation}
Here the ellipsis refer to non-$\alpha$ terms in the expansion of $\bf \langle
v\times b\rangle$ (see Eq.(4)).
Thus for isotropic situations $\alpha={1\over 3}[-\langle{\bf v.(\nabla\times
v)}\rangle+\langle{\bf b.(\nabla\times b)}\rangle]$. Thus $\alpha$ is
proportional to the difference in the fluid and magnetic helicities, a result
we have already obtained through a more detailed calculation above. In our
model fluid helicity is statistically constant in time, but magnetic helicity
grows in time and hence $\alpha\rightarrow 0$ in the long time limit.

\section{summary}
\label{summ}
Thus in conclusions we have shown how the initial exponential growth of 
the magnetic fields in a turbulent dynamo can be arrested by the action of the
magnetic fields on the velocity fields. Our mechanism required that the parity
breaking parts of the velocity and magnetic field variances must have the same
sign, which must be the case in any physical system. 
It is also worth noting the role of the symmetries of the velocity and magnetic
field correlation functions. The antisymetric part, which is also  an
irreducible part of the $\langle v_i v_j\rangle$ tensor is responsible
for the growth. Again the antisymetric (irreducible) part, 
part of the $\langle b_ib_j\rangle$ tensor is responsible 
for stabilisation.  Even though our explicit
calculations were done by using simple initial conditions for the
calculational convenience, the results that we obtain are general
enough and it is obvious that the feedback mechanism is independent
of the details of the intial conditions. Thus our results should be 
valid for more realistic initial conditions also. We have also demonstrated
that our results can be easily understood within the first-order smoothing
approximation. From the point of view of nonlinear
systems, our results can be interpreted as `non-linear stabilisation
of linear instabilities', qualitatively similar to the well-known example of
the Kuramato-Shivanisky (KS) equation in one and two dimensions. 
This is linearly
unstable for small wavenumbers, but the nonlinear term stabilises it. In fact,
the long wavelength properties of the KS equation is same as that of the
stochastically driven Kardar-Parisi-Zhang equation (KPZ)
\cite{rahul} in one and two
dimensions. However we must add words of caution while making the comparison
between the KS-KPZ problem and the dynamo problem. In the KS-KPZ case the KS
equation is not stochastically driven. The long wavelength instability serves
as drive. However, in the present case of dynamo, the velocity field (or the
NS equation) is stochastically driven. Recently it has been shown that the
stochastically and determinitically driven NS equations belong to the same
multiscaling universality class \cite{asain} (in the inertial frames). Even
though the same has not been shown for $3d$MHD, it is probably true there too.
So with some confidence we can draw the analogy between the KS-KPZ problem and
the dynamo problem which we discussed here. 
An important issue is still however left open. In fully developed $3d$MHD, in
the steady state,
correlation and response functions exhibit dynamical scaling and the {\em
dynamic exponent} $z=2/3$ \cite{four,abjkb}, which means renormalise
dissipations (kinetic as well as magnetic) diverge $\sim k^{-4/3}$ for a
wavenumber $k$ belonging to the inertial range. Even for decaying MHD with
initial K41-type correlations this turns out to be true \cite{decay} where
equal time correlations exhibit dynamical scaling with $z=2/3$. The question
is, what it is in the intial transient of dynamo growth? K41-type of spectra
suggest the roughness exponent for both the velocity and the magentic is 1/3
\cite{yakhot,abjkb,decay}. This, together with Galilean invariance for the
$3d$MHD equations give $z=2/3$. So, unless $\alpha$ coefficients too pick up
divergent corrections dynamo growth will be subdominant to dissipative decay in
the long wavelength limit. However, as our results [Eqs.(\ref{alphak41})]
suggest, the alpha coefficients diverge in the long wavelength limit,
indicating that they pick up $k$-dependent singular corrections. Simple minded
calculations \cite{abhik1} suggest $\alpha\sim k^{-1/3}$. 
If this is really the case then
neither growth nor dissipative decay dominate in the inertial range; the
sign of $(\alpha-\mu)$ determines it. However this kind of RG calculations
suffer from few technical problems \cite{yakhot}. Thus to settle it
conclusively one requires more sophesticated technique and/or numerical
simulations.  This issue of divergent effective viscosities in the inertial
range assumes importance as it may help to overcome some of the non-linear
restrictions as discussed by Vainshtein and Cattaneo \cite{vain}. 
A system of magnetohydrodynamic turbulence in a rotating
frame, after the saturation time (i.e., after which there is no net 
dynamo growth) belongs to the universality class of usual three-dimensional 
magnetohydrodynamic turbulence in a laboratory. This can be seen easily as 
both in the lab and rotating frames, the roughness and the dynamic exponents
can be calculated exactly by using the Galilean invariance and
noise-nonrenormalisation conditions \cite{yakhot,abjkb}. However, the same
cannot be immediately said about the multiscaling exponents of the higher order
structure functions. Further investigation is required in this direction. It
will also be very interesting to find out the detailed quantitative dependences
of $\alpha$ on the magnetic Prandtl number ($P_m$) in view of the recent 
results that $P_m$ is connected to other dimensionless numbers like the ratio
of the total cross helicity to the kinetic energy in the steady state
\cite{abun}.

\section{Acknowledgement}
The author wishes to thank J. K. Bhattacharjee for drawing his attention to
this problem.

\end{document}